\documentstyle[12pt]{article}
\begin{document}
\baselineskip 16pt plus 2pt minus 2pt


\begin{center}

{\large \bf   QCD ACCURATELY PREDICTS THE INDUCED PSEUDOSCALAR
COUPLING CONSTANT}

\vspace{2.5cm}

V. Bernard$^{\dag ,1}$, N. Kaiser$^{\S ,2}$, Ulf-G. Mei\ss ner$^{\dag ,3}$

\vspace{0.5cm}

$^{\dag}$Physique Th\'{e}orique, Centre de Recherches Nucl\'{e}aires et
Universit\'{e}
Louis Pasteur de Strasbourg, B.P. 20, F-67037 Strasbourg Cedex 2, France

\vspace{0.5cm}

$^{\S}$TU M\"unchen, Physik Department T30, James-Franck-Stra\ss e,
D-85747 Garching, Germany

\vspace{0.5cm}

email: $^1$bernard@crnhp4.in2p3.fr, $^2$nkaiser@physik.tu-muenchen.de,
$^3$meissner@crnvax.in2p3.fr

\end{center}

\vspace{3cm}

\begin{center}

{\bf ABSTRACT:}

\vspace{0.1cm}

Using chiral Ward identities of QCD, we derive a relation for
the induced pseudoscalar coupling constant which is accurate within a
few percent, $g_P = 8.44 \pm 0.16$.

\end{center}

\vspace{5cm}

\noindent CRN-94/13 \hfill March 1994

\vfill

\pagebreak

\vspace{0.5cm}

\noindent The structure of the nucleon
 as probed by weak charged currents is encoded in
two form factors, the axial and the induced pseudoscalar ones. While much
attention has been focused on the first, the latter is generally believed to
be understood well in terms of pion pole dominance
as indicated from ordinary muon capture experiments,
$\mu^- + p \rightarrow \nu_\mu + n$ (see e.g. ref.\cite{Ber}).
However, it now seems feasible to measure the induced pseudoscalar
coupling constant (the form factor evaluated at $t = -0.88M_\mu^2$) within
a few percent accuracy via new techniques which allow to minimize the
uncertainty in the neutron detection \cite{Taqq}. We will demonstrate here
that one is also able to calculate this fundamental quantity within a few
percent accuracy by making use of the chiral Ward identities of QCD.

\vspace{0.3cm}

\noindent To be specific,
consider the matrix--element of the isovector axial quark
current, $A_\mu^a = q \gamma_\mu \gamma_5 (\tau^a / 2) q$, between nucleon
states \cite{su2}
\begin{equation}
<N(p')|\, A_\mu^a \, |N(p)> = \bar{u}(p') \biggl[ \gamma_\mu \, G_A (t) +
\frac{(p' -p)_\mu}{2m} \, G_P(t) \biggr] \gamma_5 \frac{\tau^a}{2} u(p)
\label{e1}
\end{equation}

\noindent with $t=(p'-p)^2$ the invariant momentum transfer squared and
$m$ the nucleon mass. The form
of eq.\ref{e1} follows from  Lorentz invariance, isospin conservation and the
discrete symmetries C, P and T. $G_A (t)$ is called the nucleon
axial form factor and $G_P (t)$ the induced pseudoscalar form  factor.
Here, we are interested in the pseudoscalar coupling constant
\begin{equation}
g_P = \frac{M_\mu}{2 m} G_P(t = -0.88 M_\mu^2 )
\label{e2}
\end{equation}

\noindent as can be measured in ordinary muon capture.
Our aim is to give an accurate prediction for $g_P$ in terms of
well--known physical parameters. For doing that, we exploit the chiral Ward
identity of QCD,
\begin{equation}
\partial^\mu [ \bar{q} \gamma_\mu \gamma_5 \frac{\tau^a}{2} q ] =
\hat{m} \bar{q} i \gamma_5 \tau^a q
\label{e3}
\end{equation}

\noindent with $\hat m$ the average light quark mass \cite{iso}.
Sandwiching eq.\ref{e3}
between nucleon states, one obtains \cite{GSS}
\begin{equation}
mG_A(t) + \frac{t}{4 m} G_P(t) = 2 \hat{m} \, B \, m^0 \,
 g_A^0 \frac{1 + h(t)}{M_\pi^2 - t}
\label{e4}
\end{equation}

\noindent where the supersript '0' denotes quantities in the chiral limit,
${\cal Q}
= {\cal Q^0} [ 1 + {\cal O}(\hat m )]$. Here, $B= -<0|\bar{u}u|0>/F_\pi^2$ is
the order parameter of the spontaneous chiral symmetry breaking and
$F_\pi$ the weak pion decay constant determined from the decay $\pi^+
\rightarrow \mu^+ + \nu_\mu$.
The pion pole in eq.\ref{e4} originates
from the direct coupling of the pseudoscalar density  to the pion, $<0|\bar{q}
i \gamma_5 \tau^a q | \pi^b> = \delta^{ab} G_\pi$ \cite{GL}. The residue at the
pion pole $t = M_\pi^2$ is \cite{GSS} \cite{GL}
\begin{equation}
\hat{m} \, G_\pi \, g_{\pi N} = g_{\pi N} \, F_\pi \, M_\pi^2
\label{e5}
\end{equation}

\noindent with $g_{\pi N}$ the strong pion--nucleon coupling constant.
To go further, we make use of heavy
baryon chiral perturbation theory (HBCHPT) as detailed ref.\cite{BKKM}. To
order $q^4$, we have

\begin{eqnarray}
G_A (t) =  g_A \bigl( 1 + \frac{r_A^2}{6} t \bigr)  \label{e6a} \\
h(t) =  {\rm const}  - \frac{2 b'_{11}}{F_\pi^2} t
\label{e6b}
\end{eqnarray}

\noindent with $g_A = G_A (0)$ the axial--vector coupling constant,
$r_A^2$ the mean square axial radius of the nucleon and
$b'_{11}$ a low--energy constant \cite{const}.
The reason for the linear dependence
in eqs.\ref{e6a},\ref{e6b} is the following.
The corresponding form factors $G_A (t)$ and
$h(t)$ have a cut starting at $t = (3 M_\pi)^2$ which in the chiral expansion
first shows up at two--loop order ${\cal O}(q^5)$ ( $q$ denotes a small
external momentum or a meson mass). Therefore, the contribution to order $q^4$
must be polynomial in $t$.
Furthermore,  from chiral counting it follows that the
$t^2$-terms are related to order $q^5$ of the full matrix--elements. Putting
pieces together, we arrive at
\begin{equation}
m \, g_A + m \, g_A \frac{r_A^2}{6} t + \frac{t}{4m} G_p(t) =
\frac{g_{\pi N} F_\pi}{M_\pi^2-t} t + g_{\pi N} F_\pi + \frac{2 b'_{11} M_\pi^2
g_{\pi N}}{F_\pi}
\label{e7}
\end{equation}

\noindent where we have used $2 \hat{m} B g_A^0 m^0 = M_\pi^2 ( g_{\pi N}
F_\pi + {\cal O}(M_\pi^2) )$. At $t=0$, eq.\ref{e7} reduces to the
Goldberger--Treiman discrepancy \cite{GSS} \cite{BKKM}
\begin{equation}
 g_A \, m = g_{\pi N} \,F_\pi
\, \biggl( 1+ \frac{2 b'_{11}}{F_\pi^2} \, M_\pi^2 \biggr)
\label{e8}
\end{equation}

\noindent Eq.\ref{e8} clarifies the meaning of the low--energy constant
$b'_{11}$. Finally, $G_P (t)$ can be isolated from eq.\ref{e7},
\begin{equation}
G_P (t) = \frac{4 m g_{\pi N} F_\pi}{M_\pi^2 - t} \, - \frac{2}{3} \, g_A \,
m^2 \, r_A^2 \, + {\cal O}(t, M_\pi^2)
\label{e9}
\end{equation}

\noindent A few remarks are in order. First, notice that only physical and
well--determined parameters enter in eq.\ref{e9}. Second, while the first
term on the right--hand--side of eq.\ref{e9} is of order $q^{-2}$, the second
one is ${\cal O}(q^0)$ and the corrections not calculated are of order $q^2$.
For $g_P$, this leads to
\begin{equation}
g_P  = \frac{2 M_\mu g_{\pi N} F_\pi}{M_\pi^2 + 0.88M_\mu^2} \,
 - \frac{1}{3} \, g_A \, M_\mu \, m \, r_A^2
\label{e10}
\end{equation}

\noindent Indeed, the relation eq.\ref{e10} has been derived long time ago by
Wolfenstein \cite{WOL} using a once--subtracted dispersion relation for the
right--hand--side of eq.\ref{e4} (weak PCAC). It is gratifying that
Wolfenstein's result can be firmly based on the systematic chiral expansion of
low energy QCD Green functions. In chiral perturbation theory, one could in
principle calculate the corrections to eq.\ref{e10} by performing a two--loop
calculation while in Wolfenstein's method these could only be estimated. To
stress it again, the main ingredient to arrive at eq.\ref{e10} in HBCHPT is the
linear $t$--dependence in eqs.\ref{e6a},\ref{e6b}. Since we are interested here
in a very small momentum transfer, $t = -0.88M_\mu^2 \simeq -0.5 M_\pi^2$,
curvature terms of order $t^2$ have to be negligible. If one uses for example
the dipole parametrization for the axial form factor, $G_A (t)= (1 -
t/M_A^2)^{-2}$, the $t^2$--term amounts to a 1.3$\%$ correction to the one
linear in $t$.

\vspace{0.3cm}

\noindent The masses $m$, $M_\mu$ and $M_\pi = M_{\pi^+}$ are accurately known
and so are $F_\pi = 92.5 \pm 0.2$ MeV and $g_A = 1.2573 \pm 0.0028$ \cite{PDG}.
The situation concerning the strong pion--nucleon coupling constant is less
favourable. The methodologically best determination based on dispersion theory
gave $g_{\pi N}^2 / 4 \pi = 14.28 \pm 0.36$ \cite{LB}, more recent
determinations seem to favor smaller values \cite{AWP}.
 We use here $g_{\pi N} = 13.31 \pm
0.34$ \cite{Hoeh}. The most accurate determinations of $r_A$ stem from
(anti)neutrino--nucleon scattering, the world average being $r_A = 0.65 \pm
0.03$ fm. This uncertainty plays, however, no role in the final result since
the second term on the right--hand--side of eq.\ref{e10} is much smaller than
the first one,
\begin{equation}
g_P  = (8.89 \pm 0.23)  - (0.45 \pm 0.04) = 8.44 \pm 0.16
\label{e11}
\end{equation}

\noindent  The uncertainties in eq.\ref{e11} stem from the range of $g_{\pi N}$
and from the one for $r_A$ for the first and second term, in order. For the
final result on $g_P$, we have added these uncertainties in quadrature. A
measurement with a 2$\%$ accuracy of $g_P$ could therefore cleanly separate
between the pion pole contribution and the improved CHPT result. This would
mean a significant progress in our understanding of this fundamental
low--energy parameter since the presently available determinations have too
large error bars to disentangle these values (see e.g. \cite{Ber}). In fact,
one might turn the argument around and eventually use a precise determination
of $g_P$ to get an additional determination of the strong pion--nucleon
coupling
constant which has been at the center of much controversy over the last years.

\vspace{0.3cm}

\noindent To summarize, we have shown that the chiral Ward identities allow
to predict the induced pseudoscalar coupling constant entirely in terms of
well--determined physical parameters within a few percent accuracy. As already
noted by Wolfenstein \cite{WOL}, an accurate empirical determination of this
quantity  therefore poses a stringent test on our understanding of the
underlying dynamics which is believed to be realized in the effective
low--energy field theory of QCD (i.e. chiral perturbation theory).

\vspace{1cm}

\noindent {\bf Acknowledgments}
\bigskip

\noindent We thank D. Taqqu for interesting us in this problem and Prof. G.
H\"ohler for instructive comments.


\vspace{1cm}

\begin{thebibliography}{99}

\bibitem{Ber} J. Bernabeu, Nucl.Phys.A374(1982)593c.

\bibitem{Taqq}
D. Taqqu, contribution presented at the International Workshop on
"Large Experiments
at Low Energy Hadron Machines", PSI, Switzerland, April 1994;
and  private communication.

\bibitem{su2} Throughout, we work in flavor SU(2), i.e. $q^T = (u,d)$.

\bibitem{iso} The isopin--violating effects from the quark mass difference
$m_d - m_u$ can safely be neglected, compare the discussion in section 12
of ref.\cite{GL}.

\bibitem{GSS} J. Gasser, M.E. Sainio and A. Svarc, Nucl.Phys.B307(1988)779.

\bibitem{GL} J. Gasser and H. Leutwyler, Ann.Phys.(NY)158(1984)124.

\bibitem{BKKM} V.Bernard, N. Kaiser, J. Kambor and Ulf-G. Mei\ss ner,
Nucl. Phys. B388(1992)315.

\bibitem{const} We do not specify the constant appearing in eq.7 since it
is not needed explicitely in the following.

\bibitem{WOL} L. Wolfenstein, in: High-Energy Physics and Nuclear Structure,
ed. S. Devons (Plenum, New York, 1970) p.661.

\bibitem{PDG} Particle Data Group, Phys.Rev.D45(1993)S1.

\bibitem{LB} G. H\"ohler, in Landolt--B\"ornstein, vol. 9b2, ed. H. Schopper
(Springer,Berlin,1983).

\bibitem{AWP} For example, Arndt and collaborators have concluded from
fixed-t dispersion relations that $g^2_{\pi N} / 4 \pi = 13.72 \pm 0.15$,
which amounts to a 2$\%$ decrease of the value
for $g_{\pi N}$ from ref.\cite{LB}. See
R.A. Arndt et al., $\pi N$ Newsletter No.8(1993)37. Other determinations of
$g_{\pi N}$ like e.g. from $NN$--scattering involve some model--dependence
and are therefore less stringent.

\bibitem{Hoeh} G. H\"ohler, private communication.

\end{thebibliography}
\end{document}